\makeatletter
\@ifundefined{@parse@version@dash}{%
\def\@parse@version#1{\@parse@version@0#1}
\def\@parse@version@#1/#2/#3#4#5\@nil{%
\@parse@version@dash#1-#2-#3#4\@nil}
\def\@parse@version@dash#1-#2-#3#4#5\@nil{%
  \if\relax#2\relax\else#1\fi#2#3#4 }
}{}
\makeatother  

\documentclass[prl,twocolumn,superscriptaddress,showpacs,floatfix,longbibliography]{revtex4-2}
\usepackage{mathrsfs,braket}
\usepackage{amssymb, amsbsy, amsmath, latexsym, dsfont, array, layout, graphicx,mathrsfs,color,ulem,bm}
\usepackage[colorlinks=true,citecolor=blue,urlcolor=blue,linkcolor=blue]{hyperref}

\begin{document}
\title{Quantum time reflection and refraction of ultracold atoms}

\author{Zhaoli Dong}
\thanks{These authors contributed equally to this work}
\author{Hang Li}
\thanks{These authors contributed equally to this work}
\author{Tuo Wan}
\thanks{These authors contributed equally to this work}
\affiliation{%
Interdisciplinary Center of Quantum Information, State Key Laboratory of Modern Optical Instrumentation, Zhejiang Province Key Laboratory of Quantum Technology and Device, School of Physics, Zhejiang University, Hangzhou 310027, China
}%
\author{Qian Liang}
\affiliation{%
Interdisciplinary Center of Quantum Information, State Key Laboratory of Modern Optical Instrumentation, Zhejiang Province Key Laboratory of Quantum Technology and Device, School of Physics, Zhejiang University, Hangzhou 310027, China
}%
\author{Zhaoju Yang}
\email{zhaojuyang@zju.edu.cn}
\affiliation{%
Interdisciplinary Center of Quantum Information, State Key Laboratory of Modern Optical Instrumentation, Zhejiang Province Key Laboratory of Quantum Technology and Device, School of Physics, Zhejiang University, Hangzhou 310027, China
}%
\author{Bo Yan}
\email{yanbohang@zju.edu.cn}
\affiliation{%
Interdisciplinary Center of Quantum Information, State Key Laboratory of Modern Optical Instrumentation, Zhejiang Province Key Laboratory of Quantum Technology and Device, School of Physics, Zhejiang University, Hangzhou 310027, China
}%

\date{\today}

\begin{abstract}
Time reflection and refraction are temporal analogies of the spatial boundary effects derived from Fermat’s principle. They occur when classical waves strike a time boundary where an abrupt change in the properties of the medium is introduced. The main features of time-reflected and refracted waves are the shift of frequency and conservation of momentum, which offer a new degree of freedom for steering extreme waves and controlling phases of matter. The concept was originally proposed for manipulating optical waves more than five decades ago. However, due to the extreme challenges in the ultrafast engineering of the optical materials, the experimental realization of the time boundary effects remains elusive. Here, we introduce a time boundary into a momentum lattice of ultracold atoms and simultaneously demonstrate the time reflection and refraction experimentally. Through launching a Gaussian-superposed state into the Su-Schrieffer-Heeger (SSH) atomic chain, we observe the time-reflected and refracted waves when the input state strikes a time boundary. Furthermore, we detect a transition from time reflection/refraction to localization with increasing strength of disorder and show that the time boundary effects are robust against considerable disorder. Our work opens a new avenue for future exploration of time boundaries and spatiotemporal lattices, and their interplay with non-Hermiticity and many-body interactions.
\end{abstract}

\maketitle

\begin{figure*}[!t]
	\centering
	\includegraphics[width=0.95\linewidth]{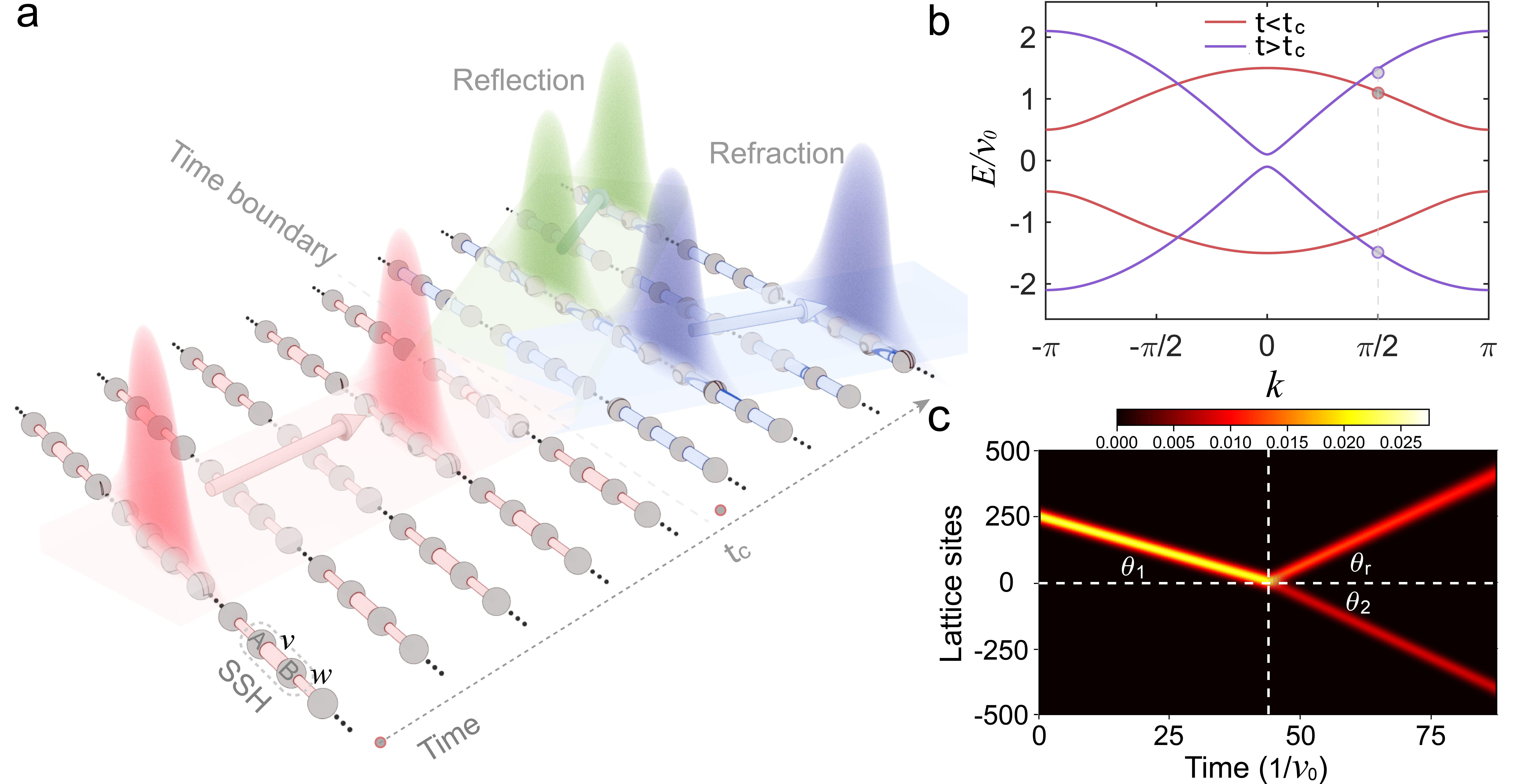}
\caption{\textbf{Schematic of time reflection and refraction.} \textbf{a.} Time reflection and refraction in the SSH model in an ultracold-atom system. The time boundary is implemented by introducing uniformly an abrupt change of the coupling amplitudes. \textbf{b.} The band structures before and after the time boundary. The coupling amplitudes are $w_0=v_0/2$ before the time boundary while $w_1=-1.1v_0$ after the time boundary. And the numerical simulation of the time boundary dynamics with $M=500$ unit cells ($N=1000$ lattice sites) is shown in  \textbf{c.} The initial Gaussian wave packet is determined by Eq. (2) with $k_0=\pi/2, m_0=-120$ and $\sigma=12.5$.}
	\label{scheme}
\end{figure*}

{\bf Introduction.}
Laws of reflection and refraction derived from Fermat’s principle have governed the classical waves striking spatial boundaries for centuries. The space-time duality suggests that there exist temporal analogs of the spatial boundary effects \cite{AKHMANOV1969,KOLNER1994,Mendonca2002,Xiao2014,PhysRevLett.120.043902} and these time boundary effects have raised burgeoning interest recently \cite{Zhou2020,Lyubarov2021,Dikopoltsev2022}. Generally, a wave propagating through a time boundary where an abrupt change in the properties of the medium occurs uniformly experiences time reflection and refraction. In stark contrast to the spatial boundary effects, the frequency of the time reflected and refracted waves is shifted and the wavevector remains unchanged, as required by the causality relationship which unravels waves can not move backward in time. Direct manipulations of the time boundaries may lead to many opportunities, such as photonic time crystals \cite{Lustig:18,PhysRevLett.126.163902,Lyubarov2021,Sharabi:22,PhysRevLett.128.186802} and Floquet metamaterials \cite{Yin2022, Weitenberg2021}. Despite the intriguing directions, the observation of time boundary effects requires a large and rapid enough change in the refractive index, which is extremely challenging in optical materials \cite{Lustig:21,twoleg2209}. Until recently, the time refraction has been observed at the optical frequency \cite{Zhou2020,Lustig:21} and the time reflection has only been reported in the classical water-wave \cite{Bacot2016} and transmission-line systems \cite{Moussa2022}. The lack of experimental realization leaves the investigation of various temporal boundary effects largely unexplored.

On the other hand, a quantum platform based on ultracold atoms is of great importance to modern quantum technology, with applications ranging from quantum simulation to quantum computation \cite{2008_RMPcold,2012_qugas,Gross2017,Schafer2020}. In the past decade, another boundary effect- the topological boundary effect has drawn a lot of attention and has been explored with ultracold atoms \cite{Zhai2015,Goldman2016,Cooper2019}. For example, the Haldane model and quantum Hall effect were experimentally realized in ultracold atom systems \cite{Jotzu2014,Mancini2015,Stuhl2015} and the chiral edge currents were detected by exploiting the synthetic dimension in trapped cold atoms \cite{Mancini2015,Stuhl2015}. The SSH model supporting topological interface states has been reported with ultracold atoms in both optical superlattice \cite{Folling2007,Atala2013} and momentum-space lattice \cite{Meier2016,2019_npj}. Moving from Hermitian to non-Hermitian \cite{Yao2018, Ding2022} systems, our group has reported the non-Hermitian skin effect by utilizing a lossy SSH atomic chain \cite{Liang2022,Gou2020}. Intuitively, these advances lead us to the question of whether the aforementioned time boundary effects can be realized in a quantum system and easily controlled. The flexible controllability and excellent quantum nature of ultracold atoms provide a new avenue for addressing these issues and future exploration of various time boundary effects.

In this work, we experimentally engineer an SSH chain to study the time reflection and refraction effects in a momentum lattice of ultracold atoms \cite{2016_pra,2018_science}. With an abrupt change in the SSH chain's couplings, we can introduce a time boundary. An initial wave packet with a definite wave vector is prepared in our system and its evolution dynamics are recorded when striking the time boundary. The time-reflected and refracted wave packets are observed and well governed by the laws of time reflection and refraction. Furthermore, by introducing quasi-periodic disorder into our lattice after the time boundary, we unravel that the time reflection and refraction are robust against considerable disorder. Our results introduce the concept of time boundary into quantum gases beyond the scope of adiabatic approximation and offer opportunities for further exploring various time boundary effects such as momentum gaps and spatiotemporal optical lattices \cite{Lustig:18,PhysRevLett.126.163902,Sharabi:22}, and their interplay with non-Hermiticity \cite{Yao2018, Ding2022} and many-body interactions \cite{Abanin2019,Randall2021}.

{\bf Theoretical model.}
The schematic of our model is shown in Fig. \ref{scheme}a. We start from the SSH model with the Hamiltonian
\begin{equation}\label{eq1}
H=\sum_{n}[v\hat{a}_{n}^{\dagger}\hat{b}_{n}+w(t)\hat{a}_{n}^{\dagger}\hat{b}_{n+1}]+h.c.
\end{equation}  
where $v$ ($w$) is the coupling strength between nearest-neighbor sites,  $\hat{a}_{n}, \hat{b}_{n}$ ($\hat{a}_{n}^{\dagger}, \hat{b}_{n}^{\dagger}$) are the annihilation (creation) operators for particles at the A, B sites of the $n$th unit cell. To realize the time boundary, we set the coupling strength $v$ to be a constant $v=v_0$, and  $w(t)$ to be a time-dependent variable. It can be easily obtained that the SSH model is a two-band system and the dispersion relationship is $E_{\pm}(k)=\pm \sqrt{w^{2}+v^{2}+2wv\cos k}$, where the corresponding eigenstates are $\phi  _{\pm}(k)=\frac{1}{\sqrt{2}}[1,\pm e^{i\theta _{k}}]$, and $\tan\theta _{k}=w\sin k/(v+w\cos k)$. 
Inspired by the Ref. \cite{twoleg2209}, we set an abrupt change on the coupling strength of $w(t)$ at $t_c$ in order to introduce a time boundary into the ultracold-atom system. Figure \ref{scheme}b shows the typical band structures of the SSH model before ($w(t)=w_0=v_0/2$, red curves) and after ($w(t)=w_1=-1.1v_0$, purple curves) the time boundary.

Before the time boundary with $t<t_c$, an initial wavepacket that is a superposition of the eigenstates around the red dot in Fig. \ref{scheme}b corresponding to the wave vector of $k_{0}$ is excited. The wavepacket will propagate to the right with a positive group velocity of $(\partial E_+/\partial k)_{k=k_0}$. After the time boundary with $t>t_c$, the band structure is abruptly switched to the purple curves and the initial state will be projected to the new eigenstates. Because the abrupt change of the coupling strength at the time boundary is spatially uniform, the momentum is a conserved quantity. Therefore, at $t=t_c$, the initial state will split into two propagating wave packets corresponding to opposite energies (indicated by the purple dots in Fig. \ref{scheme}b) and group velocities with the same amplitude but opposite signs. The wave packet with the same sign of the group velocity as the initial state is called the time-refracted wave, while the one with the opposite sign of the group velocity is called the time-reflected wave, as depicted in Fig. \ref{scheme}a.

\begin{figure*}[t]
	\centering
	\includegraphics[width=1.0\linewidth]{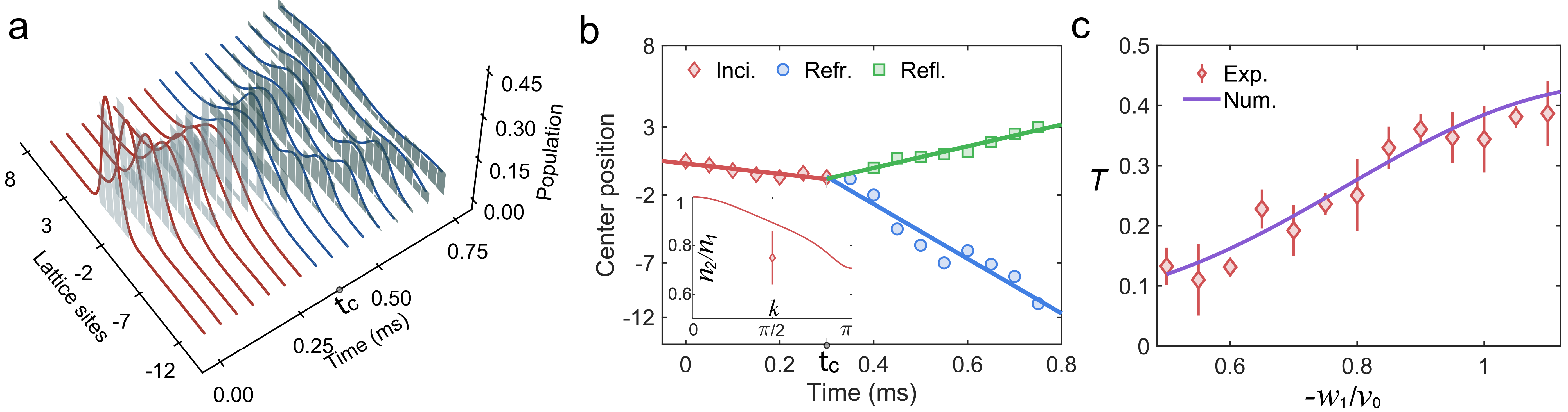}
	\caption{\textbf{Experimental observation of time reflection and refraction.} \textbf{a.} Time boundary effects in the momentum lattice. The initial superposed state is $\Psi_0=0.12e^{(-i\pi/2)}|-2\rangle+0.38|-1\rangle+0.38|0\rangle+0.12e^{(i\pi/2)}|1\rangle$ \cite{supp}. The red and blue curves are the Double Gaussian fitting before and after the time boundary. \textbf{b.} Central positions of the original incident wave (red dashed curve), time-reflected (green dashed line), and time-refracted (blue dashed line) waves. Insert line shows $n_2/n_1$ with different $k$, and the dot shows the experimental data for $k=\pi/2$. \textbf{c.} The ratio of time refraction as a function of the coupling rate $w_1/v_0$. The red dots indicate the experimental results. The purple curve represents the numerical result based on the four-site initial Gaussian wave packet.}
        \label{figure2}
\end{figure*}

Based on the above model, we show the numerical results of the time boundary effects in Fig. \ref{scheme}c. The initial Gaussian wave packet obeys 
 \begin{equation}\label{eq2}
\left| \psi ^{m}_{0}(k_{0}) \right>=\frac{1}{\sqrt{\sigma }\sqrt[4]{\pi }}e^{-(m-m_{0})^{2}/2\sigma^{2}}e^{ik_{0}m}\left| \phi  _{+}(k_{0}) \right>, 
\end{equation}
where $\left| \psi ^{m}_{0}(k_{0}) \right>$ stands for the initial state of $m$th unit cell, $m_{0}$ is the central cell number of the wave packet and $\sigma$ represents the width of the wave packet. As we can see clearly, the initial wave packet first propagates to the right, strikes at the time boundary, and then splits into the time-refracted and time-reflected waves with opposite-signed group velocities, which agrees well with the above discussion.

In our quantum system, laws of the time reflection and refraction are not as straightforward as those in optics \cite{Mendonca2002,Plansinis2015, Zhou2020}, since there exists no well-defined velocity or refractive index for quantum gases in the time dimension. Fortunately, with the help of the group velocity defined by $(\partial E_+/\partial k)_{k=k_0}$, we can introduce the effective refractive index and generalize the laws of the time refraction as (see more details in \cite{supp}):
\begin{equation}\label{eq3}
{n_2}/{n_1}={\cos\theta_2}/{\cos\theta_1}, 
\end{equation}
where $n_1$ ($n_2$) is the effective refractive index for $t<t_c$ ($t>t_c$), $\theta_{1,2}$ are the incidence and refraction angles defined in Fig.\ref{scheme}c. Therefore, the effective refractive index for the time boundary effects can be retrieved from the above equation and $n_2/n_1=0.89$ for $k=\pi/2$. For the time reflection in our system, it is easy to obtain that the reflection angle is $\theta_{r}=\theta_2$ due to the symmetric band structure with respect to zero energy. Details of the derivations are shown in \cite{supp}.

{\bf Experimental observation of time boundary effects.} 
In experiments, we start with a BEC of about $2 \times 10^{5}$ $ ^{87}\textrm{Rb}$ atoms in a  dipole trap~\cite{2018_JOSAB,2021_npj,ABcage2022}. Multiple Bragg laser pairs ($\lambda$= 1064 nm) couple the 21 discrete momentum states (labeled as $\left| n \right>$ ($n \in  \mathbb{Z} $)) to form the SSH chain with the typical coupling rate in the energy scale of $k$Hz. In our experiment, the time boundary is realized by changing the Bragg lasers with the Acoustic-optical modulators, whose speed can reach as fast as MHz, which is much larger than the energy scale ($\sim k$Hz) of the system. Therefore, it ensures that the time boundary effect can be observed in our ultracold atom system. More details about the experiment can be found in \cite{supp}. 

To observe the time boundary effects, an initial superposed state according to Eq.(\ref{eq2}) with $k_0=\pi/2$ \cite{supp} is launched on four sites at the center of the momentum lattice. The coupling amplitudes are choosen $v_0=3.47(3)$ kHz, $w_0=1.75(1)$ kHz. The time evolution of the initial state is shown as red curves in Fig.~\ref{figure2}a. At the time of $t_c$=0.33 ms, an abrupt change of the coupling to $w_1=-3.50(2)$ kHz imposes a boundary in the time dimension. At $t>t_c$, the evolution of the waves is shown as blue curves in Fig.~\ref{figure2}a. The wave packet is clearly split into two parts: a time-refracted wave and a time-reflected wave. To further characterize the evolution features, we use the double-Gaussian fitting to extract the central positions from the envelopes of the wave packets. Figure \ref{figure2}b shows the center positions of the incident, time-reflected, and time-refracted waves as a function of time. The effective index is extracted to be $n_2/n_1=(0.75\pm0.11)$, which is close to the theoretic calculation as shown in the insert of Fig.~\ref{figure2}b. Note that the deviations from the numerical results stem from the four-site excitation and finite-size effect \cite{supp}. The central observation of these experiments is that the initial wave packet propagates to the right, strikes the time boundary, and then divides into time-refracted and time-reflected waves, which provides a direct observation of the time boundary effects.

With the experimental observations, the time refraction ratio can be easily calculated \cite{supp}. We next sweep $w_1/v_0$ and the ratio of the time refraction as a function of $w_1/v_0$ is shown in Fig. \ref{figure2}c. The time refraction ratio varies from about 0.1 to 0.4 when $w_1/v_0$ is within the accessible range of $[-0.5, -1.1]$ in the experiment. The experimental result agrees well with numerical simulations \cite{supp}. Along this vein, the ratio of the time reflection is extracted by $1-T$ and within the range from 0.6 to 0.9. The triumph of the time reflection in our system is because of the relatively large amplitude of the projection probability of the upper eigenstate before $t_c$ and the one after $t_c$. This distinct feature of our model makes it easier for observing time reflections than in optical materials. Note that the total time reflection of the incident wave can be obtained when we change the Hamiltonian by a negative sign \cite{supp}, which is dubbed 'Loschmidt echo' that is nowadays a standard method in quantum control \cite{Raitzsch2008, PhysRevLett.124.160603}.

\begin{figure}[!t]
\centering
\includegraphics[width=0.75\linewidth]{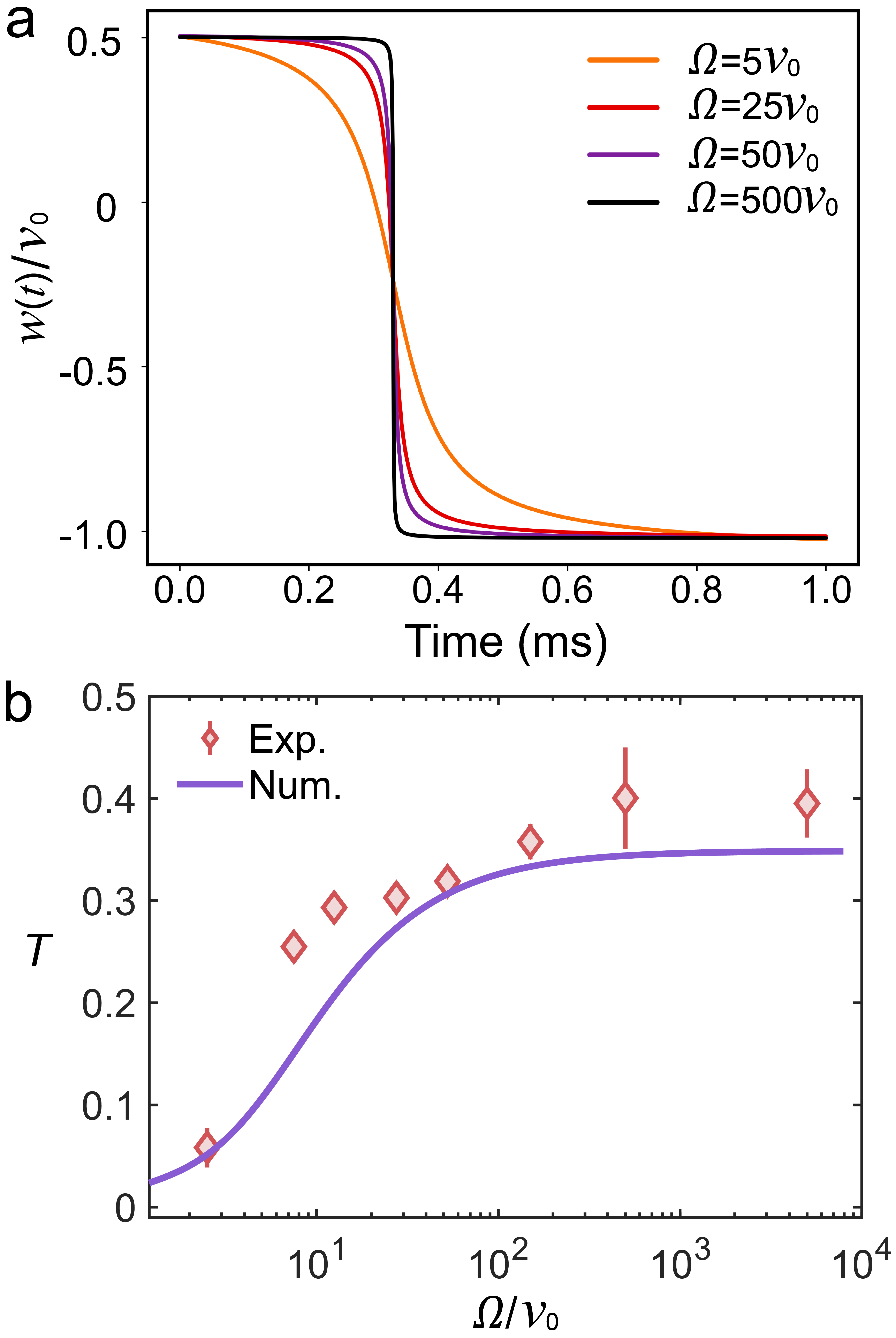}
\caption{\textbf{Varying time boundary.} \textbf{a.} Different varying waveforms of $w(t)$ with the parameters $\delta=$3 and $\alpha=0.5$. \textbf{b.} Time refraction ratio as a function of $\Omega$. The red dots indicate the experimental results. The purple curve represents the result from the numerical simulation.}
\label{figure3}
\end{figure}

{\bf Varying time boundary.} 
In the above experiments, the abrupt change of the coupling strength is a step function. If $w(t)$ is changed slowly enough, the band structure of the system will adiabatically be deformed from the red to blue lines in Fig. \ref{scheme}b, which includes an opposite change of the group velocity. Consequently, only the time reflection will be left. Naturally, this raises the question of what if the time boundary doesn't change so fast or so slowly? 

To investigate this issue, we adopt different smooth time boundaries with an analytical form of $w(t)=\left [  -({\delta }/{\pi})\arctan(\Omega (t-t_c))-\alpha\right ]w_{0}$, where the parameter $\delta$ and $\alpha$ are constants to ensure the changing function has a definite beginning and end values, the parameter $\Omega$ controls the sharpness of the waveform. As shown in Fig. \ref{figure3}a, $w(t)$ smoothly changed from $w_0/v_0$=0.5 to $w_1/v_0=-1$ with different speeds. Smaller $\Omega/v_0$ indicates a slower varying waveform of $w(t)$, and the infinite $\Omega/v_0$ corresponds to the step function. 

By sweeping the parameter of $\Omega/v_0$ within the range from 2.5 to 5000, the time refraction ratio $T$ is quantitatively controlled, as shown in Fig.~\ref{figure3}b. When the time boundary is sharp enough ($\Omega/v_0>100$), the time refraction ratio stays at the stable plateau of $T$=0.35, which indicates that the time boundary effects observed in Fig.~\ref{figure2} are robust against considerable imperfection of the time boundary. When the $\Omega$ is small enough ($\Omega/v_0<10$), the time-refracted wave, as well as the time-reflected wave, disappears in the experiments, which verifies the fact that the time boundary effects require a fast enough switching rate of the time boundary.

\begin{figure}[!t]
\centering
\includegraphics[width=0.7\linewidth]{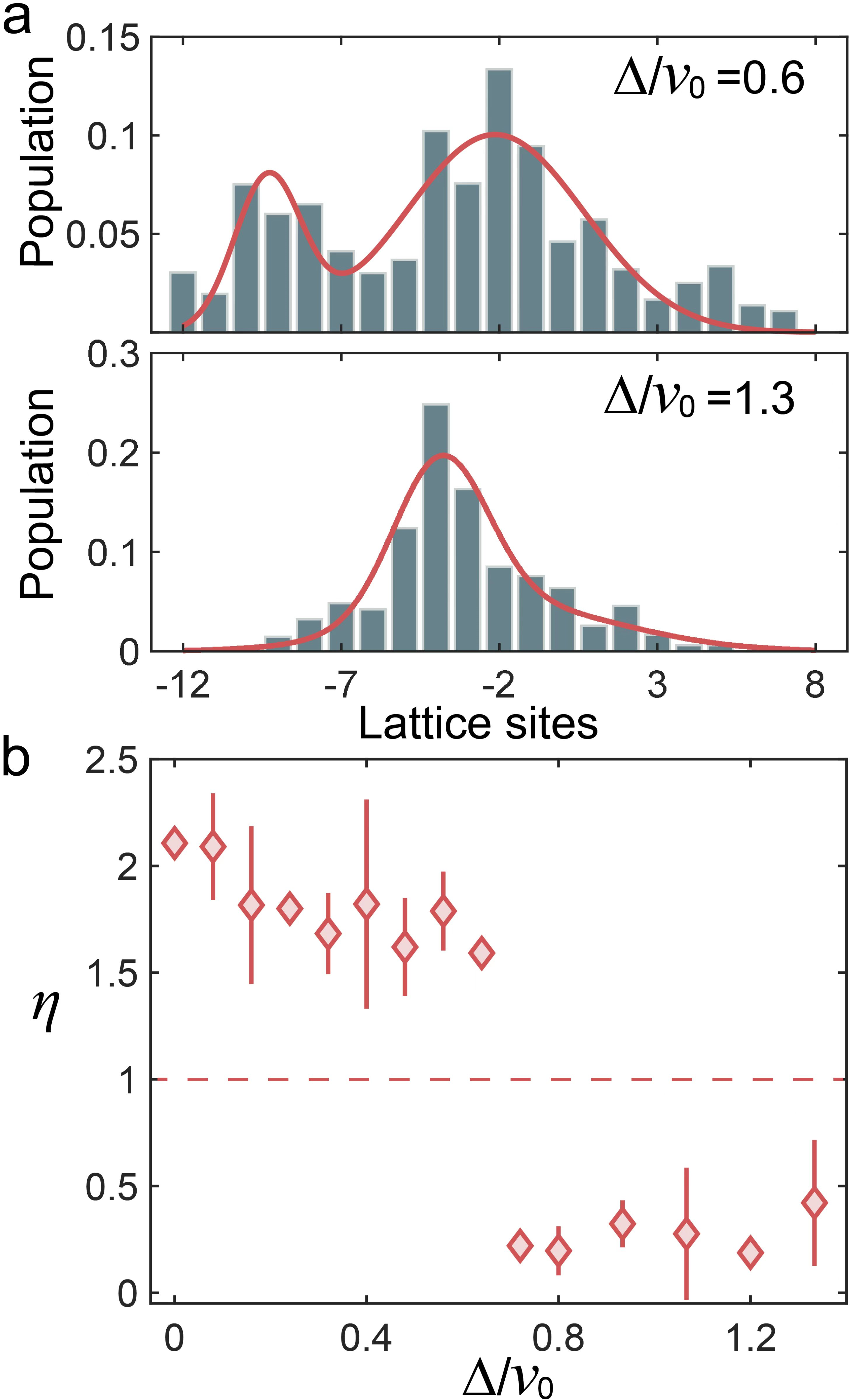}
\caption{\textbf{Robustness of the time boundary effect.} \textbf{a.} Population distributions measured at $t$=0.7 ms with disorder strength of $\Delta/v_0=0.6$ (upper panel) and $\Delta/v_0=1.3$ (lower panel), respectively. $\varphi=5\pi/8$ and the other parameters are the same as Fig.~\ref{figure2}a. The solid lines are fitted lines with function $P_d=A_{1}\exp [ -(n-n_{1})^{2}/2\sigma _{1}^{2}]+A_{2}\exp[ -(n-n_{1}-\delta n)^{2}/2\sigma _{2}^{2}]$. \textbf{b.} The ratio of distance-to-width as a function of disorder strength of $\Delta/v_0$. The critical point $\eta=1$ indicates the transition point where the double-peak curve changes into a single-peak curve.}
\label{figure4}
\end{figure}

{\bf Robustness of time boundary effects.} 
Having shown the time boundary effects can survive imperfection of the time boundary, we expect that these effects are also robust against disorder. To unveil the interplay between disorder and the time boundary effects, we experimentally engineer the on-site quasiperiodic disorder in our system \cite{An2021,PhysRevLett.123.070405,PhysRevB.91.014108,PhysRevB.87.134202}, the additional Hamitonian of disorder is written as $H_\text{dis}=-(\Delta_{n}^{(a)} \hat{a}_{n}^{\dagger} \hat{a}_{n}+\Delta_{n}^{(b)} \hat{b}_{n}^{\dagger}\hat{b}_{n})$, where $\Delta_{n}^{(i)}=\Delta\cos[2\pi\alpha (2n+n_{0})+\varphi]$, $n_{0}=0$ (when $i=a$) or $n_{0}=1$ (when $i=b$), $\Delta$, $\varphi$ are the amplitude and phase of quasiperiodicity, and the irrational number $\alpha=(\sqrt{5}-1)/4$. As shown in Fig. \ref{figure4}a, two evolution dynamics of the time boundary effects are displayed with different disorder strengths. The time-reflected and refracted waves (two peaks) still survive under the relatively high disorder of $\Delta/v_0=0.6$, which shows strong evidence of the robustness of time boundary effects. If the disorder strength is large enough, the system will eventually enter into the Anderson localization and only one peak can be found in the lower panel of Fig.~\ref{figure4}a, which means the time boundary effects disappear. 

To further characterize the robustness of the time boundary effects, we have fitted the measured distribution using a double-Gaussian function. The results are shown in the red curves of Fig. \ref{figure4}a. Specifically, we have defined a distance-to-width ratio $\eta=\left|\Delta n\right|/\sqrt{\sigma _{1}^{2}+\sigma _{2}^{2}}$ to describe this transition. The measured $\eta$ values with different disorder strengths are shown in Fig. \ref{figure4}b, which indicates that the double-peak curve can be preserved with a considerable range of the disorder strength. The transition from double to single peak appears at $\Delta/v_0\simeq0.7$, which is consistent with the numerical result of about 0.8 (more details see \cite{supp}).

{\bf Conclusion.}
We have experimentally demonstrated the time boundary effects in a momentum lattice of ultracold atoms. The time reflection and refraction of ultracold atoms have been observed simultaneously and manipulated by controlling the coupling amplitudes and sharpness of the time boundary. Furthermore, we have shown that the time boundary effects are robust against considerable disorder. Our results pave the way for the future exploration of the time and spatiotemporal crystals \cite{Lustig:18, Sharabi:22} in the ultracold momentum-lattice system, especially for the topics about the amplification in the momentum gaps \cite{Lyubarov2022}, and their interplay with topological phases \cite{Lustig:18}, non-Hermiticity \cite{Yao2018, Ding2022} and many-body interactions \cite{Heugel2019,Randall2021}.

\begin{acknowledgments}
{\it Acknowledgement:}
We acknowledge the support from the National Key Research and Development Program of China under Grant No. 2022YFA1404203, and the National Natural Science Foundation of China under Grant Nos. U21A20437, 12074337, and 12174339, Zhejiang Provincial Natural Science Foundation of China under Grant No. LR21A040002 and LR23A040003, Zhejiang Provincial Plan for Science and Technology Grant No. 2020C01019, the Fundamental Research Funds for the Central Universities under Grant No. 2021FZZX001-02 and the Excellent Youth Science Foundation Project (Overseas).
\end{acknowledgments}

\bibliographystyle{apsrev4-2}
\bibliography{main}

\begin{thebibliography}{53}%
\makeatletter
\providecommand \@ifxundefined [1]{%
 \@ifx{#1\undefined}
}%
\providecommand \@ifnum [1]{%
 \ifnum #1\expandafter \@firstoftwo
 \else \expandafter \@secondoftwo
 \fi
}%
\providecommand \@ifx [1]{%
 \ifx #1\expandafter \@firstoftwo
 \else \expandafter \@secondoftwo
 \fi
}%
\providecommand \natexlab [1]{#1}%
\providecommand \enquote  [1]{``#1''}%
\providecommand \bibnamefont  [1]{#1}%
\providecommand \bibfnamefont [1]{#1}%
\providecommand \citenamefont [1]{#1}%
\providecommand \href@noop [0]{\@secondoftwo}%
\providecommand \href [0]{\begingroup \@sanitize@url \@href}%
\providecommand \@href[1]{\@@startlink{#1}\@@href}%
\providecommand \@@href[1]{\endgroup#1\@@endlink}%
\providecommand \@sanitize@url [0]{\catcode `\\12\catcode `\$12\catcode
  `\&12\catcode `\#12\catcode `\^12\catcode `\_12\catcode `\%12\relax}%
\providecommand \@@startlink[1]{}%
\providecommand \@@endlink[0]{}%
\providecommand \url  [0]{\begingroup\@sanitize@url \@url }%
\providecommand \@url [1]{\endgroup\@href {#1}{\urlprefix }}%
\providecommand \urlprefix  [0]{URL }%
\providecommand \Eprint [0]{\href }%
\providecommand \doibase [0]{https://doi.org/}%
\providecommand \selectlanguage [0]{\@gobble}%
\providecommand \bibinfo  [0]{\@secondoftwo}%
\providecommand \bibfield  [0]{\@secondoftwo}%
\providecommand \translation [1]{[#1]}%
\providecommand \BibitemOpen [0]{}%
\providecommand \bibitemStop [0]{}%
\providecommand \bibitemNoStop [0]{.\EOS\space}%
\providecommand \EOS [0]{\spacefactor3000\relax}%
\providecommand \BibitemShut  [1]{\csname bibitem#1\endcsname}%
\let\auto@bib@innerbib\@empty
\bibitem [{\citenamefont {S.~Akhmanov}\ and\ \citenamefont
  {Chirkin}(1969)}]{AKHMANOV1969}%
  \BibitemOpen
  \bibfield  {author} {\bibinfo {author} {\bibfnamefont {A.~S.}\ \bibnamefont
  {S.~Akhmanov}}\ and\ \bibinfo {author} {\bibfnamefont {A.}~\bibnamefont
  {Chirkin}},\ }\href {http://www.jetp.ras.ru/cgi-bin/dn/e_028_04_0748.pdf}
  {\bibfield  {journal} {\bibinfo  {journal} {Sov. Phys. JETP}\ }\textbf
  {\bibinfo {volume} {28}},\ \bibinfo {pages} {748} (\bibinfo {year}
  {1969})}\BibitemShut {NoStop}%
\bibitem [{\citenamefont {Kolner}(1994)}]{KOLNER1994}%
  \BibitemOpen
  \bibfield  {author} {\bibinfo {author} {\bibfnamefont {B.}~\bibnamefont
  {Kolner}},\ }\href
  {https://ieeexplore.ieee.org/document/301659/citations#citations} {\bibfield
  {journal} {\bibinfo  {journal} {IEEE Journal of Quantum Electronics}\
  }\textbf {\bibinfo {volume} {30}},\ \bibinfo {pages} {1951} (\bibinfo {year}
  {1994})}\BibitemShut {NoStop}%
\bibitem [{\citenamefont {Mendonca}\ and\ \citenamefont
  {Shukla}(2002)}]{Mendonca2002}%
  \BibitemOpen
  \bibfield  {author} {\bibinfo {author} {\bibfnamefont {J.~T.}\ \bibnamefont
  {Mendonca}}\ and\ \bibinfo {author} {\bibfnamefont {P.~K.}\ \bibnamefont
  {Shukla}},\ }\href {https://doi.org/10.1238/Physica.Regular.065a00160}
  {\bibfield  {journal} {\bibinfo  {journal} {Physica Scripta}\ }\textbf
  {\bibinfo {volume} {65}},\ \bibinfo {pages} {160} (\bibinfo {year}
  {2002})}\BibitemShut {NoStop}%
\bibitem [{\citenamefont {Xiao}\ \emph {et~al.}(2014)\citenamefont {Xiao},
  \citenamefont {Maywar},\ and\ \citenamefont {Agrawal}}]{Xiao2014}%
  \BibitemOpen
  \bibfield  {author} {\bibinfo {author} {\bibfnamefont {Y.~Z.}\ \bibnamefont
  {Xiao}}, \bibinfo {author} {\bibfnamefont {D.~N.}\ \bibnamefont {Maywar}},\
  and\ \bibinfo {author} {\bibfnamefont {G.~P.}\ \bibnamefont {Agrawal}},\
  }\href {https://doi.org/10.1364/OL.39.000574} {\bibfield  {journal} {\bibinfo
   {journal} {Optics Letters}\ }\textbf {\bibinfo {volume} {39}},\ \bibinfo
  {pages} {574} (\bibinfo {year} {2014})}\BibitemShut {NoStop}%
\bibitem [{\citenamefont {Vezzoli}\ \emph {et~al.}(2018)\citenamefont
  {Vezzoli}, \citenamefont {Bruno}, \citenamefont {DeVault}, \citenamefont
  {Roger}, \citenamefont {Shalaev}, \citenamefont {Boltasseva}, \citenamefont
  {Ferrera}, \citenamefont {Clerici}, \citenamefont {Dubietis},\ and\
  \citenamefont {Faccio}}]{PhysRevLett.120.043902}%
  \BibitemOpen
  \bibfield  {author} {\bibinfo {author} {\bibfnamefont {S.}~\bibnamefont
  {Vezzoli}}, \bibinfo {author} {\bibfnamefont {V.}~\bibnamefont {Bruno}},
  \bibinfo {author} {\bibfnamefont {C.}~\bibnamefont {DeVault}}, \bibinfo
  {author} {\bibfnamefont {T.}~\bibnamefont {Roger}}, \bibinfo {author}
  {\bibfnamefont {V.~M.}\ \bibnamefont {Shalaev}}, \bibinfo {author}
  {\bibfnamefont {A.}~\bibnamefont {Boltasseva}}, \bibinfo {author}
  {\bibfnamefont {M.}~\bibnamefont {Ferrera}}, \bibinfo {author} {\bibfnamefont
  {M.}~\bibnamefont {Clerici}}, \bibinfo {author} {\bibfnamefont
  {A.}~\bibnamefont {Dubietis}},\ and\ \bibinfo {author} {\bibfnamefont
  {D.}~\bibnamefont {Faccio}},\ }\href
  {https://doi.org/10.1103/PhysRevLett.120.043902} {\bibfield  {journal}
  {\bibinfo  {journal} {Phys. Rev. Lett.}\ }\textbf {\bibinfo {volume} {120}},\
  \bibinfo {pages} {043902} (\bibinfo {year} {2018})}\BibitemShut {NoStop}%
\bibitem [{\citenamefont {Zhou}\ \emph {et~al.}(2020)\citenamefont {Zhou},
  \citenamefont {Alam}, \citenamefont {Karimi}, \citenamefont {Upham},
  \citenamefont {Reshef}, \citenamefont {Liu}, \citenamefont {Willner},\ and\
  \citenamefont {Boyd}}]{Zhou2020}%
  \BibitemOpen
  \bibfield  {author} {\bibinfo {author} {\bibfnamefont {Y.}~\bibnamefont
  {Zhou}}, \bibinfo {author} {\bibfnamefont {M.~Z.}\ \bibnamefont {Alam}},
  \bibinfo {author} {\bibfnamefont {M.}~\bibnamefont {Karimi}}, \bibinfo
  {author} {\bibfnamefont {J.}~\bibnamefont {Upham}}, \bibinfo {author}
  {\bibfnamefont {O.}~\bibnamefont {Reshef}}, \bibinfo {author} {\bibfnamefont
  {C.}~\bibnamefont {Liu}}, \bibinfo {author} {\bibfnamefont {A.~E.}\
  \bibnamefont {Willner}},\ and\ \bibinfo {author} {\bibfnamefont {R.~W.}\
  \bibnamefont {Boyd}},\ }\href {https://doi.org/10.1038/s41467-020-15682-2}
  {\bibfield  {journal} {\bibinfo  {journal} {Nature Communications}\ }\textbf
  {\bibinfo {volume} {11}},\ \bibinfo {pages} {2180} (\bibinfo {year}
  {2020})}\BibitemShut {NoStop}%
\bibitem [{\citenamefont {Lyubarov}\ \emph {et~al.}(2021)\citenamefont
  {Lyubarov}, \citenamefont {Lumer}, \citenamefont {Dikopoltsev}, \citenamefont
  {Lustig}, \citenamefont {Sharabi},\ and\ \citenamefont
  {Segev}}]{Lyubarov2021}%
  \BibitemOpen
  \bibfield  {author} {\bibinfo {author} {\bibfnamefont {M.}~\bibnamefont
  {Lyubarov}}, \bibinfo {author} {\bibfnamefont {Y.}~\bibnamefont {Lumer}},
  \bibinfo {author} {\bibfnamefont {A.}~\bibnamefont {Dikopoltsev}}, \bibinfo
  {author} {\bibfnamefont {E.}~\bibnamefont {Lustig}}, \bibinfo {author}
  {\bibfnamefont {Y.}~\bibnamefont {Sharabi}},\ and\ \bibinfo {author}
  {\bibfnamefont {M.}~\bibnamefont {Segev}},\ }\href@noop {} {\bibfield
  {journal} {\bibinfo  {journal} {2021 Conference on Lasers and Electro-optics
  (cleo)}\ ,\ \bibinfo {pages} {IEEE}} (\bibinfo {year} {2021})}\BibitemShut
  {NoStop}%
\bibitem [{\citenamefont {Dikopoltsev}\ \emph {et~al.}(2022)\citenamefont
  {Dikopoltsev}, \citenamefont {Sharabi}, \citenamefont {Lyubarov},
  \citenamefont {Lumer}, \citenamefont {Tsesses}, \citenamefont {Lustig},
  \citenamefont {Kaminer},\ and\ \citenamefont {Segev}}]{Dikopoltsev2022}%
  \BibitemOpen
  \bibfield  {author} {\bibinfo {author} {\bibfnamefont {A.}~\bibnamefont
  {Dikopoltsev}}, \bibinfo {author} {\bibfnamefont {Y.}~\bibnamefont
  {Sharabi}}, \bibinfo {author} {\bibfnamefont {M.}~\bibnamefont {Lyubarov}},
  \bibinfo {author} {\bibfnamefont {Y.}~\bibnamefont {Lumer}}, \bibinfo
  {author} {\bibfnamefont {S.}~\bibnamefont {Tsesses}}, \bibinfo {author}
  {\bibfnamefont {E.}~\bibnamefont {Lustig}}, \bibinfo {author} {\bibfnamefont
  {I.}~\bibnamefont {Kaminer}},\ and\ \bibinfo {author} {\bibfnamefont
  {M.}~\bibnamefont {Segev}},\ }\href {https://doi.org/10.1073/pnas.2119705119}
  {\bibfield  {journal} {\bibinfo  {journal} {Proceedings of the National
  Academy of Sciences of the United States of America}\ }\textbf {\bibinfo
  {volume} {119}},\ \bibinfo {pages} {e2119705119} (\bibinfo {year}
  {2022})}\BibitemShut {NoStop}%
\bibitem [{\citenamefont {Lustig}\ \emph {et~al.}(2018)\citenamefont {Lustig},
  \citenamefont {Sharabi},\ and\ \citenamefont {Segev}}]{Lustig:18}%
  \BibitemOpen
  \bibfield  {author} {\bibinfo {author} {\bibfnamefont {E.}~\bibnamefont
  {Lustig}}, \bibinfo {author} {\bibfnamefont {Y.}~\bibnamefont {Sharabi}},\
  and\ \bibinfo {author} {\bibfnamefont {M.}~\bibnamefont {Segev}},\ }\href
  {https://doi.org/10.1364/OPTICA.5.001390} {\bibfield  {journal} {\bibinfo
  {journal} {Optica}\ }\textbf {\bibinfo {volume} {5}},\ \bibinfo {pages}
  {1390} (\bibinfo {year} {2018})}\BibitemShut {NoStop}%
\bibitem [{\citenamefont {Sharabi}\ \emph {et~al.}(2021)\citenamefont
  {Sharabi}, \citenamefont {Lustig},\ and\ \citenamefont
  {Segev}}]{PhysRevLett.126.163902}%
  \BibitemOpen
  \bibfield  {author} {\bibinfo {author} {\bibfnamefont {Y.}~\bibnamefont
  {Sharabi}}, \bibinfo {author} {\bibfnamefont {E.}~\bibnamefont {Lustig}},\
  and\ \bibinfo {author} {\bibfnamefont {M.}~\bibnamefont {Segev}},\ }\href
  {https://doi.org/10.1103/PhysRevLett.126.163902} {\bibfield  {journal}
  {\bibinfo  {journal} {Phys. Rev. Lett.}\ }\textbf {\bibinfo {volume} {126}},\
  \bibinfo {pages} {163902} (\bibinfo {year} {2021})}\BibitemShut {NoStop}%
\bibitem [{\citenamefont {Sharabi}\ \emph {et~al.}(2022)\citenamefont
  {Sharabi}, \citenamefont {Dikopoltsev}, \citenamefont {Lustig}, \citenamefont
  {Lumer},\ and\ \citenamefont {Segev}}]{Sharabi:22}%
  \BibitemOpen
  \bibfield  {author} {\bibinfo {author} {\bibfnamefont {Y.}~\bibnamefont
  {Sharabi}}, \bibinfo {author} {\bibfnamefont {A.}~\bibnamefont
  {Dikopoltsev}}, \bibinfo {author} {\bibfnamefont {E.}~\bibnamefont {Lustig}},
  \bibinfo {author} {\bibfnamefont {Y.}~\bibnamefont {Lumer}},\ and\ \bibinfo
  {author} {\bibfnamefont {M.}~\bibnamefont {Segev}},\ }\href
  {https://doi.org/10.1364/OPTICA.455672} {\bibfield  {journal} {\bibinfo
  {journal} {Optica}\ }\textbf {\bibinfo {volume} {9}},\ \bibinfo {pages} {585}
  (\bibinfo {year} {2022})}\BibitemShut {NoStop}%
\bibitem [{\citenamefont {Peng}(2022)}]{PhysRevLett.128.186802}%
  \BibitemOpen
  \bibfield  {author} {\bibinfo {author} {\bibfnamefont {Y.}~\bibnamefont
  {Peng}},\ }\href {https://doi.org/10.1103/PhysRevLett.128.186802} {\bibfield
  {journal} {\bibinfo  {journal} {Phys. Rev. Lett.}\ }\textbf {\bibinfo
  {volume} {128}},\ \bibinfo {pages} {186802} (\bibinfo {year}
  {2022})}\BibitemShut {NoStop}%
\bibitem [{\citenamefont {Yin}\ \emph {et~al.}(2022)\citenamefont {Yin},
  \citenamefont {Galiffi},\ and\ \citenamefont {Al{\`u}}}]{Yin2022}%
  \BibitemOpen
  \bibfield  {author} {\bibinfo {author} {\bibfnamefont {S.}~\bibnamefont
  {Yin}}, \bibinfo {author} {\bibfnamefont {E.}~\bibnamefont {Galiffi}},\ and\
  \bibinfo {author} {\bibfnamefont {A.}~\bibnamefont {Al{\`u}}},\ }\href
  {https://doi.org/10.1186/s43593-022-00015-1} {\bibfield  {journal} {\bibinfo
  {journal} {eLight}\ }\textbf {\bibinfo {volume} {2}},\ \bibinfo {pages} {8}
  (\bibinfo {year} {2022})}\BibitemShut {NoStop}%
\bibitem [{\citenamefont {Weitenberg}\ and\ \citenamefont
  {Simonet}(2021)}]{Weitenberg2021}%
  \BibitemOpen
  \bibfield  {author} {\bibinfo {author} {\bibfnamefont {C.}~\bibnamefont
  {Weitenberg}}\ and\ \bibinfo {author} {\bibfnamefont {J.}~\bibnamefont
  {Simonet}},\ }\href {https://doi.org/10.1038/s41567-021-01316-x} {\bibfield
  {journal} {\bibinfo  {journal} {Nature Physics}\ }\textbf {\bibinfo {volume}
  {17}},\ \bibinfo {pages} {1342} (\bibinfo {year} {2021})}\BibitemShut
  {NoStop}%
\bibitem [{\citenamefont {Lustig}\ \emph {et~al.}(2021)\citenamefont {Lustig},
  \citenamefont {Saha}, \citenamefont {Bordo}, \citenamefont {DeVault},
  \citenamefont {Chowdhury}, \citenamefont {Sharabi}, \citenamefont
  {Boltasseva}, \citenamefont {Cohen}, \citenamefont {Shalaev},\ and\
  \citenamefont {Segev}}]{Lustig:21}%
  \BibitemOpen
  \bibfield  {author} {\bibinfo {author} {\bibfnamefont {E.}~\bibnamefont
  {Lustig}}, \bibinfo {author} {\bibfnamefont {S.}~\bibnamefont {Saha}},
  \bibinfo {author} {\bibfnamefont {E.}~\bibnamefont {Bordo}}, \bibinfo
  {author} {\bibfnamefont {C.}~\bibnamefont {DeVault}}, \bibinfo {author}
  {\bibfnamefont {S.~N.}\ \bibnamefont {Chowdhury}}, \bibinfo {author}
  {\bibfnamefont {Y.}~\bibnamefont {Sharabi}}, \bibinfo {author} {\bibfnamefont
  {A.}~\bibnamefont {Boltasseva}}, \bibinfo {author} {\bibfnamefont
  {O.}~\bibnamefont {Cohen}}, \bibinfo {author} {\bibfnamefont {V.~M.}\
  \bibnamefont {Shalaev}},\ and\ \bibinfo {author} {\bibfnamefont
  {M.}~\bibnamefont {Segev}},\ }in\ \href
  {https://doi.org/10.1364/CLEO_QELS.2021.FF2H.1} {\emph {\bibinfo {booktitle}
  {Conference on Lasers and Electro-Optics}}}\ (\bibinfo  {publisher} {Optica
  Publishing Group},\ \bibinfo {year} {2021})\ p.\ \bibinfo {pages}
  {FF2H.1}\BibitemShut {NoStop}%
\bibitem [{\citenamefont {Olivia Y.~Long}\ and\ \citenamefont
  {Fan}(2022)}]{twoleg2209}%
  \BibitemOpen
  \bibfield  {author} {\bibinfo {author} {\bibfnamefont {A.~D.}\ \bibnamefont
  {Olivia Y.~Long}, \bibfnamefont {Kai~Wang}}\ and\ \bibinfo {author}
  {\bibfnamefont {S.}~\bibnamefont {Fan}},\ }\href
  {https://arxiv.org/abs/2209.03539} {\bibfield  {journal} {\bibinfo  {journal}
  {arXiv:2209.03539v2}\ } (\bibinfo {year} {2022})}\BibitemShut {NoStop}%
\bibitem [{\citenamefont {Bacot}\ \emph {et~al.}(2016)\citenamefont {Bacot},
  \citenamefont {Labousse}, \citenamefont {Eddi}, \citenamefont {Fink},\ and\
  \citenamefont {Fort}}]{Bacot2016}%
  \BibitemOpen
  \bibfield  {author} {\bibinfo {author} {\bibfnamefont {V.}~\bibnamefont
  {Bacot}}, \bibinfo {author} {\bibfnamefont {M.}~\bibnamefont {Labousse}},
  \bibinfo {author} {\bibfnamefont {A.}~\bibnamefont {Eddi}}, \bibinfo {author}
  {\bibfnamefont {M.}~\bibnamefont {Fink}},\ and\ \bibinfo {author}
  {\bibfnamefont {E.}~\bibnamefont {Fort}},\ }\href
  {https://doi.org/10.1038/nphys3810} {\bibfield  {journal} {\bibinfo
  {journal} {Nature Physics}\ }\textbf {\bibinfo {volume} {12}},\ \bibinfo
  {pages} {972} (\bibinfo {year} {2016})}\BibitemShut {NoStop}%
\bibitem [{\citenamefont {Moussa}\ \emph {et~al.}(2022)\citenamefont {Moussa},
  \citenamefont {Xu}, \citenamefont {Yin}, \citenamefont {Galiffi},
  \citenamefont {Radi},\ and\ \citenamefont {Alù}}]{Moussa2022}%
  \BibitemOpen
  \bibfield  {author} {\bibinfo {author} {\bibfnamefont {H.}~\bibnamefont
  {Moussa}}, \bibinfo {author} {\bibfnamefont {G.}~\bibnamefont {Xu}}, \bibinfo
  {author} {\bibfnamefont {S.}~\bibnamefont {Yin}}, \bibinfo {author}
  {\bibfnamefont {E.}~\bibnamefont {Galiffi}}, \bibinfo {author} {\bibfnamefont
  {Y.}~\bibnamefont {Radi}},\ and\ \bibinfo {author} {\bibfnamefont
  {A.}~\bibnamefont {Alù}},\ }\href@noop {} {\bibfield  {journal} {\bibinfo
  {journal} {arXiv:2208.07236}\ } (\bibinfo {year} {2022})}\BibitemShut
  {NoStop}%
\bibitem [{\citenamefont {Bloch}\ \emph {et~al.}(2008)\citenamefont {Bloch},
  \citenamefont {Dalibard},\ and\ \citenamefont {Zwerger}}]{2008_RMPcold}%
  \BibitemOpen
  \bibfield  {author} {\bibinfo {author} {\bibfnamefont {I.}~\bibnamefont
  {Bloch}}, \bibinfo {author} {\bibfnamefont {J.}~\bibnamefont {Dalibard}},\
  and\ \bibinfo {author} {\bibfnamefont {W.}~\bibnamefont {Zwerger}},\ }\href
  {https://doi.org/10.1103/RevModPhys.80.885} {\bibfield  {journal} {\bibinfo
  {journal} {Reviews of Modern Physics}\ }\textbf {\bibinfo {volume} {80}},\
  \bibinfo {pages} {885} (\bibinfo {year} {2008})}\BibitemShut {NoStop}%
\bibitem [{\citenamefont {Bloch}\ \emph {et~al.}(2012)\citenamefont {Bloch},
  \citenamefont {Dalibard},\ and\ \citenamefont {Nascimbene}}]{2012_qugas}%
  \BibitemOpen
  \bibfield  {author} {\bibinfo {author} {\bibfnamefont {I.}~\bibnamefont
  {Bloch}}, \bibinfo {author} {\bibfnamefont {J.}~\bibnamefont {Dalibard}},\
  and\ \bibinfo {author} {\bibfnamefont {S.}~\bibnamefont {Nascimbene}},\
  }\href {https://doi.org/10.1038/NPHYS2259} {\bibfield  {journal} {\bibinfo
  {journal} {Nature Physics}\ }\textbf {\bibinfo {volume} {8}},\ \bibinfo
  {pages} {267} (\bibinfo {year} {2012})}\BibitemShut {NoStop}%
\bibitem [{\citenamefont {Gross}\ and\ \citenamefont
  {Bloch}(2017)}]{Gross2017}%
  \BibitemOpen
  \bibfield  {author} {\bibinfo {author} {\bibfnamefont {C.}~\bibnamefont
  {Gross}}\ and\ \bibinfo {author} {\bibfnamefont {I.}~\bibnamefont {Bloch}},\
  }\href {https://doi.org/10.1126/science.aal3837} {\bibfield  {journal}
  {\bibinfo  {journal} {Science}\ }\textbf {\bibinfo {volume} {357}},\ \bibinfo
  {pages} {995} (\bibinfo {year} {2017})}\BibitemShut {NoStop}%
\bibitem [{\citenamefont {Schafer}\ \emph {et~al.}(2020)\citenamefont
  {Schafer}, \citenamefont {Fukuhara}, \citenamefont {Sugawa}, \citenamefont
  {Takasu},\ and\ \citenamefont {Takahashi}}]{Schafer2020}%
  \BibitemOpen
  \bibfield  {author} {\bibinfo {author} {\bibfnamefont {F.}~\bibnamefont
  {Schafer}}, \bibinfo {author} {\bibfnamefont {T.}~\bibnamefont {Fukuhara}},
  \bibinfo {author} {\bibfnamefont {S.}~\bibnamefont {Sugawa}}, \bibinfo
  {author} {\bibfnamefont {Y.}~\bibnamefont {Takasu}},\ and\ \bibinfo {author}
  {\bibfnamefont {Y.}~\bibnamefont {Takahashi}},\ }\href
  {https://doi.org/10.1038/s42254-020-0195-3} {\bibfield  {journal} {\bibinfo
  {journal} {Nature Reviews Physics}\ }\textbf {\bibinfo {volume} {2}},\
  \bibinfo {pages} {411} (\bibinfo {year} {2020})}\BibitemShut {NoStop}%
\bibitem [{\citenamefont {Zhai}(2015)}]{Zhai2015}%
  \BibitemOpen
  \bibfield  {author} {\bibinfo {author} {\bibfnamefont {H.}~\bibnamefont
  {Zhai}},\ }\href {https://doi.org/10.1088/0034-4885/78/2/026001} {\bibfield
  {journal} {\bibinfo  {journal} {Reports on Progress in Physics}\ }\textbf
  {\bibinfo {volume} {78}},\ \bibinfo {pages} {026001} (\bibinfo {year}
  {2015})}\BibitemShut {NoStop}%
\bibitem [{\citenamefont {Goldman}\ \emph {et~al.}(2016)\citenamefont
  {Goldman}, \citenamefont {Budich},\ and\ \citenamefont
  {Zoller}}]{Goldman2016}%
  \BibitemOpen
  \bibfield  {author} {\bibinfo {author} {\bibfnamefont {N.}~\bibnamefont
  {Goldman}}, \bibinfo {author} {\bibfnamefont {J.~C.}\ \bibnamefont
  {Budich}},\ and\ \bibinfo {author} {\bibfnamefont {P.}~\bibnamefont
  {Zoller}},\ }\href {https://doi.org/10.1038/NPHYS3803} {\bibfield  {journal}
  {\bibinfo  {journal} {Nature Physics}\ }\textbf {\bibinfo {volume} {12}},\
  \bibinfo {pages} {639} (\bibinfo {year} {2016})}\BibitemShut {NoStop}%
\bibitem [{\citenamefont {Cooper}\ \emph {et~al.}(2019)\citenamefont {Cooper},
  \citenamefont {Dalibard},\ and\ \citenamefont {Spielman}}]{Cooper2019}%
  \BibitemOpen
  \bibfield  {author} {\bibinfo {author} {\bibfnamefont {N.~R.}\ \bibnamefont
  {Cooper}}, \bibinfo {author} {\bibfnamefont {J.}~\bibnamefont {Dalibard}},\
  and\ \bibinfo {author} {\bibfnamefont {I.~B.}\ \bibnamefont {Spielman}},\
  }\href {https://doi.org/10.1103/RevModPhys.91.015005} {\bibfield  {journal}
  {\bibinfo  {journal} {Reviews of Modern Physics}\ }\textbf {\bibinfo {volume}
  {91}},\ \bibinfo {pages} {015005} (\bibinfo {year} {2019})}\BibitemShut
  {NoStop}%
\bibitem [{\citenamefont {Jotzu}\ \emph {et~al.}(2014)\citenamefont {Jotzu},
  \citenamefont {Messer}, \citenamefont {Desbuquois}, \citenamefont {Lebrat},
  \citenamefont {Uehlinger}, \citenamefont {Greif},\ and\ \citenamefont
  {Esslinger}}]{Jotzu2014}%
  \BibitemOpen
  \bibfield  {author} {\bibinfo {author} {\bibfnamefont {G.}~\bibnamefont
  {Jotzu}}, \bibinfo {author} {\bibfnamefont {M.}~\bibnamefont {Messer}},
  \bibinfo {author} {\bibfnamefont {R.}~\bibnamefont {Desbuquois}}, \bibinfo
  {author} {\bibfnamefont {M.}~\bibnamefont {Lebrat}}, \bibinfo {author}
  {\bibfnamefont {T.}~\bibnamefont {Uehlinger}}, \bibinfo {author}
  {\bibfnamefont {D.}~\bibnamefont {Greif}},\ and\ \bibinfo {author}
  {\bibfnamefont {T.}~\bibnamefont {Esslinger}},\ }\href
  {https://doi.org/10.1038/nature13915} {\bibfield  {journal} {\bibinfo
  {journal} {Nature}\ }\textbf {\bibinfo {volume} {515}},\ \bibinfo {pages}
  {237} (\bibinfo {year} {2014})}\BibitemShut {NoStop}%
\bibitem [{\citenamefont {Mancini}\ \emph {et~al.}(2015)\citenamefont
  {Mancini}, \citenamefont {Pagano}, \citenamefont {Cappellini}, \citenamefont
  {Livi}, \citenamefont {Rider}, \citenamefont {Catani}, \citenamefont {Sias},
  \citenamefont {Zoller}, \citenamefont {Inguscio}, \citenamefont {Dalmonte},\
  and\ \citenamefont {Fallani}}]{Mancini2015}%
  \BibitemOpen
  \bibfield  {author} {\bibinfo {author} {\bibfnamefont {M.}~\bibnamefont
  {Mancini}}, \bibinfo {author} {\bibfnamefont {G.}~\bibnamefont {Pagano}},
  \bibinfo {author} {\bibfnamefont {G.}~\bibnamefont {Cappellini}}, \bibinfo
  {author} {\bibfnamefont {L.}~\bibnamefont {Livi}}, \bibinfo {author}
  {\bibfnamefont {M.}~\bibnamefont {Rider}}, \bibinfo {author} {\bibfnamefont
  {J.}~\bibnamefont {Catani}}, \bibinfo {author} {\bibfnamefont
  {C.}~\bibnamefont {Sias}}, \bibinfo {author} {\bibfnamefont {P.}~\bibnamefont
  {Zoller}}, \bibinfo {author} {\bibfnamefont {M.}~\bibnamefont {Inguscio}},
  \bibinfo {author} {\bibfnamefont {M.}~\bibnamefont {Dalmonte}},\ and\
  \bibinfo {author} {\bibfnamefont {L.}~\bibnamefont {Fallani}},\ }\href
  {https://doi.org/10.1126/science.aaa8736} {\bibfield  {journal} {\bibinfo
  {journal} {Science}\ }\textbf {\bibinfo {volume} {349}},\ \bibinfo {pages}
  {1510} (\bibinfo {year} {2015})}\BibitemShut {NoStop}%
\bibitem [{\citenamefont {Stuhl}\ \emph {et~al.}(2015)\citenamefont {Stuhl},
  \citenamefont {Lu}, \citenamefont {Aycock}, \citenamefont {Genkina},\ and\
  \citenamefont {Spielman}}]{Stuhl2015}%
  \BibitemOpen
  \bibfield  {author} {\bibinfo {author} {\bibfnamefont {B.~K.}\ \bibnamefont
  {Stuhl}}, \bibinfo {author} {\bibfnamefont {H.~I.}\ \bibnamefont {Lu}},
  \bibinfo {author} {\bibfnamefont {L.~M.}\ \bibnamefont {Aycock}}, \bibinfo
  {author} {\bibfnamefont {D.}~\bibnamefont {Genkina}},\ and\ \bibinfo {author}
  {\bibfnamefont {I.~B.}\ \bibnamefont {Spielman}},\ }\href
  {https://doi.org/10.1126/science.aaa8515} {\bibfield  {journal} {\bibinfo
  {journal} {Science}\ }\textbf {\bibinfo {volume} {349}},\ \bibinfo {pages}
  {1514} (\bibinfo {year} {2015})}\BibitemShut {NoStop}%
\bibitem [{\citenamefont {Folling}\ \emph {et~al.}(2007)\citenamefont
  {Folling}, \citenamefont {Trotzky}, \citenamefont {Cheinet}, \citenamefont
  {Feld}, \citenamefont {Saers}, \citenamefont {Widera}, \citenamefont
  {Muller},\ and\ \citenamefont {Bloch}}]{Folling2007}%
  \BibitemOpen
  \bibfield  {author} {\bibinfo {author} {\bibfnamefont {S.}~\bibnamefont
  {Folling}}, \bibinfo {author} {\bibfnamefont {S.}~\bibnamefont {Trotzky}},
  \bibinfo {author} {\bibfnamefont {P.}~\bibnamefont {Cheinet}}, \bibinfo
  {author} {\bibfnamefont {M.}~\bibnamefont {Feld}}, \bibinfo {author}
  {\bibfnamefont {R.}~\bibnamefont {Saers}}, \bibinfo {author} {\bibfnamefont
  {A.}~\bibnamefont {Widera}}, \bibinfo {author} {\bibfnamefont
  {T.}~\bibnamefont {Muller}},\ and\ \bibinfo {author} {\bibfnamefont
  {I.}~\bibnamefont {Bloch}},\ }\href {https://doi.org/10.1038/nature06112}
  {\bibfield  {journal} {\bibinfo  {journal} {Nature}\ }\textbf {\bibinfo
  {volume} {448}},\ \bibinfo {pages} {1029} (\bibinfo {year}
  {2007})}\BibitemShut {NoStop}%
\bibitem [{\citenamefont {Atala}\ \emph {et~al.}(2013)\citenamefont {Atala},
  \citenamefont {Aidelsburger}, \citenamefont {Barreiro}, \citenamefont
  {Abanin}, \citenamefont {Kitagawa}, \citenamefont {Demler},\ and\
  \citenamefont {Bloch}}]{Atala2013}%
  \BibitemOpen
  \bibfield  {author} {\bibinfo {author} {\bibfnamefont {M.}~\bibnamefont
  {Atala}}, \bibinfo {author} {\bibfnamefont {M.}~\bibnamefont {Aidelsburger}},
  \bibinfo {author} {\bibfnamefont {J.~T.}\ \bibnamefont {Barreiro}}, \bibinfo
  {author} {\bibfnamefont {D.}~\bibnamefont {Abanin}}, \bibinfo {author}
  {\bibfnamefont {T.}~\bibnamefont {Kitagawa}}, \bibinfo {author}
  {\bibfnamefont {E.}~\bibnamefont {Demler}},\ and\ \bibinfo {author}
  {\bibfnamefont {I.}~\bibnamefont {Bloch}},\ }\href
  {https://doi.org/10.1038/NPHYS2790} {\bibfield  {journal} {\bibinfo
  {journal} {Nature Physics}\ }\textbf {\bibinfo {volume} {9}},\ \bibinfo
  {pages} {795} (\bibinfo {year} {2013})}\BibitemShut {NoStop}%
\bibitem [{\citenamefont {Meier}\ \emph
  {et~al.}(2016{\natexlab{a}})\citenamefont {Meier}, \citenamefont {An},\ and\
  \citenamefont {Gadway}}]{Meier2016}%
  \BibitemOpen
  \bibfield  {author} {\bibinfo {author} {\bibfnamefont {E.~J.}\ \bibnamefont
  {Meier}}, \bibinfo {author} {\bibfnamefont {F.~A.}\ \bibnamefont {An}},\ and\
  \bibinfo {author} {\bibfnamefont {B.}~\bibnamefont {Gadway}},\ }\href
  {https://doi.org/10.1038/ncomms13986} {\bibfield  {journal} {\bibinfo
  {journal} {Nature Communications}\ }\textbf {\bibinfo {volume} {7}},\
  \bibinfo {pages} {13986} (\bibinfo {year} {2016}{\natexlab{a}})}\BibitemShut
  {NoStop}%
\bibitem [{\citenamefont {Xie}\ \emph {et~al.}(2019)\citenamefont {Xie},
  \citenamefont {Gou}, \citenamefont {Xiao}, \citenamefont {Gadway},\ and\
  \citenamefont {Yan}}]{2019_npj}%
  \BibitemOpen
  \bibfield  {author} {\bibinfo {author} {\bibfnamefont {D.}~\bibnamefont
  {Xie}}, \bibinfo {author} {\bibfnamefont {W.}~\bibnamefont {Gou}}, \bibinfo
  {author} {\bibfnamefont {T.}~\bibnamefont {Xiao}}, \bibinfo {author}
  {\bibfnamefont {B.}~\bibnamefont {Gadway}},\ and\ \bibinfo {author}
  {\bibfnamefont {B.}~\bibnamefont {Yan}},\ }\href@noop {} {\bibfield
  {journal} {\bibinfo  {journal} {npj Quan. Inf.}\ }\textbf {\bibinfo {volume}
  {{5}}},\ \bibinfo {pages} {55} (\bibinfo {year} {{2019}})}\BibitemShut
  {NoStop}%
\bibitem [{\citenamefont {Yao}\ and\ \citenamefont {Wang}(2018)}]{Yao2018}%
  \BibitemOpen
  \bibfield  {author} {\bibinfo {author} {\bibfnamefont {S.}~\bibnamefont
  {Yao}}\ and\ \bibinfo {author} {\bibfnamefont {Z.}~\bibnamefont {Wang}},\
  }\href {https://doi.org/10.1103/PhysRevLett.121.086803} {\bibfield  {journal}
  {\bibinfo  {journal} {Phys. Rev. Lett.}\ }\textbf {\bibinfo {volume} {121}},\
  \bibinfo {pages} {086803} (\bibinfo {year} {2018})}\BibitemShut {NoStop}%
\bibitem [{\citenamefont {Ding}\ \emph {et~al.}(2022)\citenamefont {Ding},
  \citenamefont {Fang},\ and\ \citenamefont {Ma}}]{Ding2022}%
  \BibitemOpen
  \bibfield  {author} {\bibinfo {author} {\bibfnamefont {K.}~\bibnamefont
  {Ding}}, \bibinfo {author} {\bibfnamefont {C.}~\bibnamefont {Fang}},\ and\
  \bibinfo {author} {\bibfnamefont {G.}~\bibnamefont {Ma}},\ }\href
  {https://doi.org/10.1038/s42254-022-00516-5} {\bibfield  {journal} {\bibinfo
  {journal} {Nature Reviews Physics}\ }\textbf {\bibinfo {volume} {4}},\
  \bibinfo {pages} {745} (\bibinfo {year} {2022})}\BibitemShut {NoStop}%
\bibitem [{\citenamefont {Liang}\ \emph {et~al.}(2022)\citenamefont {Liang},
  \citenamefont {Xie}, \citenamefont {Dong}, \citenamefont {Li}, \citenamefont
  {Li}, \citenamefont {Gadway}, \citenamefont {Yi},\ and\ \citenamefont
  {Yan}}]{Liang2022}%
  \BibitemOpen
  \bibfield  {author} {\bibinfo {author} {\bibfnamefont {Q.}~\bibnamefont
  {Liang}}, \bibinfo {author} {\bibfnamefont {D.}~\bibnamefont {Xie}}, \bibinfo
  {author} {\bibfnamefont {Z.}~\bibnamefont {Dong}}, \bibinfo {author}
  {\bibfnamefont {H.}~\bibnamefont {Li}}, \bibinfo {author} {\bibfnamefont
  {H.}~\bibnamefont {Li}}, \bibinfo {author} {\bibfnamefont {B.}~\bibnamefont
  {Gadway}}, \bibinfo {author} {\bibfnamefont {W.}~\bibnamefont {Yi}},\ and\
  \bibinfo {author} {\bibfnamefont {B.}~\bibnamefont {Yan}},\ }\href
  {https://doi.org/10.1103/PhysRevLett.129.070401} {\bibfield  {journal}
  {\bibinfo  {journal} {Phys. Rev. Lett.}\ }\textbf {\bibinfo {volume} {129}},\
  \bibinfo {pages} {070401} (\bibinfo {year} {2022})}\BibitemShut {NoStop}%
\bibitem [{\citenamefont {Gou}\ \emph {et~al.}(2020)\citenamefont {Gou},
  \citenamefont {Chen}, \citenamefont {Xie}, \citenamefont {Xiao},
  \citenamefont {Deng}, \citenamefont {Gadway}, \citenamefont {Yi},\ and\
  \citenamefont {Yan}}]{Gou2020}%
  \BibitemOpen
  \bibfield  {author} {\bibinfo {author} {\bibfnamefont {W.}~\bibnamefont
  {Gou}}, \bibinfo {author} {\bibfnamefont {T.}~\bibnamefont {Chen}}, \bibinfo
  {author} {\bibfnamefont {D.}~\bibnamefont {Xie}}, \bibinfo {author}
  {\bibfnamefont {T.}~\bibnamefont {Xiao}}, \bibinfo {author} {\bibfnamefont
  {T.-S.}\ \bibnamefont {Deng}}, \bibinfo {author} {\bibfnamefont
  {B.}~\bibnamefont {Gadway}}, \bibinfo {author} {\bibfnamefont
  {W.}~\bibnamefont {Yi}},\ and\ \bibinfo {author} {\bibfnamefont
  {B.}~\bibnamefont {Yan}},\ }\href
  {https://doi.org/10.1103/PhysRevLett.124.070402} {\bibfield  {journal}
  {\bibinfo  {journal} {Phys. Rev. Lett.}\ }\textbf {\bibinfo {volume} {124}},\
  \bibinfo {pages} {070402} (\bibinfo {year} {2020})}\BibitemShut {NoStop}%
\bibitem [{\citenamefont {Meier}\ \emph
  {et~al.}(2016{\natexlab{b}})\citenamefont {Meier}, \citenamefont {An},\ and\
  \citenamefont {Gadway}}]{2016_pra}%
  \BibitemOpen
  \bibfield  {author} {\bibinfo {author} {\bibfnamefont {E.~J.}\ \bibnamefont
  {Meier}}, \bibinfo {author} {\bibfnamefont {F.~A.}\ \bibnamefont {An}},\ and\
  \bibinfo {author} {\bibfnamefont {B.}~\bibnamefont {Gadway}},\ }\href
  {https://doi.org/10.1103/PhysRevA.93.051602} {\bibfield  {journal} {\bibinfo
  {journal} {Phys. Rev. A}\ }\textbf {\bibinfo {volume} {93}},\ \bibinfo
  {pages} {051602} (\bibinfo {year} {2016}{\natexlab{b}})}\BibitemShut
  {NoStop}%
\bibitem [{\citenamefont {Meier}\ \emph {et~al.}(2018)\citenamefont {Meier},
  \citenamefont {An}, \citenamefont {Dauphin}, \citenamefont {Maffei},
  \citenamefont {Massignan}, \citenamefont {Hughes},\ and\ \citenamefont
  {Gadway}}]{2018_science}%
  \BibitemOpen
  \bibfield  {author} {\bibinfo {author} {\bibfnamefont {E.~J.}\ \bibnamefont
  {Meier}}, \bibinfo {author} {\bibfnamefont {F.~A.}\ \bibnamefont {An}},
  \bibinfo {author} {\bibfnamefont {A.}~\bibnamefont {Dauphin}}, \bibinfo
  {author} {\bibfnamefont {M.}~\bibnamefont {Maffei}}, \bibinfo {author}
  {\bibfnamefont {P.}~\bibnamefont {Massignan}}, \bibinfo {author}
  {\bibfnamefont {T.~L.}\ \bibnamefont {Hughes}},\ and\ \bibinfo {author}
  {\bibfnamefont {B.}~\bibnamefont {Gadway}},\ }\href
  {https://doi.org/10.1126/science.aat3406} {\bibfield  {journal} {\bibinfo
  {journal} {Science}\ }\textbf {\bibinfo {volume} {362}},\ \bibinfo {pages}
  {929} (\bibinfo {year} {2018})}\BibitemShut {NoStop}%
\bibitem [{\citenamefont {Abanin}\ \emph {et~al.}(2019)\citenamefont {Abanin},
  \citenamefont {Altman}, \citenamefont {Bloch},\ and\ \citenamefont
  {Serbyn}}]{Abanin2019}%
  \BibitemOpen
  \bibfield  {author} {\bibinfo {author} {\bibfnamefont {D.~A.}\ \bibnamefont
  {Abanin}}, \bibinfo {author} {\bibfnamefont {E.}~\bibnamefont {Altman}},
  \bibinfo {author} {\bibfnamefont {I.}~\bibnamefont {Bloch}},\ and\ \bibinfo
  {author} {\bibfnamefont {M.}~\bibnamefont {Serbyn}},\ }\href
  {https://doi.org/10.1103/RevModPhys.91.021001} {\bibfield  {journal}
  {\bibinfo  {journal} {Rev. Mod. Phys.}\ }\textbf {\bibinfo {volume} {91}},\
  \bibinfo {pages} {021001} (\bibinfo {year} {2019})}\BibitemShut {NoStop}%
\bibitem [{\citenamefont {Randall}\ \emph {et~al.}(2021)\citenamefont
  {Randall}, \citenamefont {Bradley}, \citenamefont {van~der Gronden},
  \citenamefont {Galicia}, \citenamefont {Abobeih}, \citenamefont {Markham},
  \citenamefont {Twitchen}, \citenamefont {Machado}, \citenamefont {Yao},\ and\
  \citenamefont {Taminiau}}]{Randall2021}%
  \BibitemOpen
  \bibfield  {author} {\bibinfo {author} {\bibfnamefont {J.}~\bibnamefont
  {Randall}}, \bibinfo {author} {\bibfnamefont {C.~E.}\ \bibnamefont
  {Bradley}}, \bibinfo {author} {\bibfnamefont {F.~V.}\ \bibnamefont {van~der
  Gronden}}, \bibinfo {author} {\bibfnamefont {A.}~\bibnamefont {Galicia}},
  \bibinfo {author} {\bibfnamefont {M.~H.}\ \bibnamefont {Abobeih}}, \bibinfo
  {author} {\bibfnamefont {M.}~\bibnamefont {Markham}}, \bibinfo {author}
  {\bibfnamefont {D.~J.}\ \bibnamefont {Twitchen}}, \bibinfo {author}
  {\bibfnamefont {F.}~\bibnamefont {Machado}}, \bibinfo {author} {\bibfnamefont
  {N.~Y.}\ \bibnamefont {Yao}},\ and\ \bibinfo {author} {\bibfnamefont {T.~H.}\
  \bibnamefont {Taminiau}},\ }\href {https://doi.org/10.1126/science.abk0603}
  {\bibfield  {journal} {\bibinfo  {journal} {Science}\ }\textbf {\bibinfo
  {volume} {374}},\ \bibinfo {pages} {abk0603} (\bibinfo {year}
  {2021})}\BibitemShut {NoStop}%
\bibitem [{sup()}]{supp}%
  \BibitemOpen
  \href@noop {} {\bibinfo  {journal} {See Supplemental Information for
  details}\ }\BibitemShut {NoStop}%
\bibitem [{\citenamefont {Plansinis}\ \emph {et~al.}(2015)\citenamefont
  {Plansinis}, \citenamefont {Donaldson},\ and\ \citenamefont
  {Agrawal}}]{Plansinis2015}%
  \BibitemOpen
\bibfield  {journal} {  }\bibfield  {author} {\bibinfo {author} {\bibfnamefont
  {B.~W.}\ \bibnamefont {Plansinis}}, \bibinfo {author} {\bibfnamefont {W.~R.}\
  \bibnamefont {Donaldson}},\ and\ \bibinfo {author} {\bibfnamefont {G.~P.}\
  \bibnamefont {Agrawal}},\ }\href
  {https://doi.org/10.1103/PhysRevLett.115.183901} {\bibfield  {journal}
  {\bibinfo  {journal} {Phys. Rev. Lett.}\ }\textbf {\bibinfo {volume} {115}},\
  \bibinfo {pages} {183901} (\bibinfo {year} {2015})}\BibitemShut {NoStop}%
\bibitem [{\citenamefont {Xie}\ \emph {et~al.}(2018)\citenamefont {Xie},
  \citenamefont {Wang}, \citenamefont {Gou}, \citenamefont {Bu},\ and\
  \citenamefont {Yan}}]{2018_JOSAB}%
  \BibitemOpen
  \bibfield  {author} {\bibinfo {author} {\bibfnamefont {D.}~\bibnamefont
  {Xie}}, \bibinfo {author} {\bibfnamefont {D.}~\bibnamefont {Wang}}, \bibinfo
  {author} {\bibfnamefont {W.}~\bibnamefont {Gou}}, \bibinfo {author}
  {\bibfnamefont {W.}~\bibnamefont {Bu}},\ and\ \bibinfo {author}
  {\bibfnamefont {B.}~\bibnamefont {Yan}},\ }\href
  {https://doi.org/10.1364/JOSAB.35.000500} {\bibfield  {journal} {\bibinfo
  {journal} {J Opt. Soc. Am. B}\ }\textbf {\bibinfo {volume} {35}},\ \bibinfo
  {pages} {500} (\bibinfo {year} {2018})}\BibitemShut {NoStop}%
\bibitem [{\citenamefont {Chen}\ \emph {et~al.}(2021)\citenamefont {Chen},
  \citenamefont {Gou}, \citenamefont {Xie}, \citenamefont {Xiao}, \citenamefont
  {Yi}, \citenamefont {Jing},\ and\ \citenamefont {Yan}}]{2021_npj}%
  \BibitemOpen
  \bibfield  {author} {\bibinfo {author} {\bibfnamefont {T.}~\bibnamefont
  {Chen}}, \bibinfo {author} {\bibfnamefont {W.}~\bibnamefont {Gou}}, \bibinfo
  {author} {\bibfnamefont {D.~Z.}\ \bibnamefont {Xie}}, \bibinfo {author}
  {\bibfnamefont {T.}~\bibnamefont {Xiao}}, \bibinfo {author} {\bibfnamefont
  {W.}~\bibnamefont {Yi}}, \bibinfo {author} {\bibfnamefont {J.}~\bibnamefont
  {Jing}},\ and\ \bibinfo {author} {\bibfnamefont {B.}~\bibnamefont {Yan}},\
  }\href {https://doi.org/10.1038/s41534-021-00417-y} {\bibfield  {journal}
  {\bibinfo  {journal} {Npj Quantum Information}\ }\textbf {\bibinfo {volume}
  {7}},\ \bibinfo {pages} {78} (\bibinfo {year} {2021})}\BibitemShut {NoStop}%
\bibitem [{\citenamefont {Li}\ \emph {et~al.}(2022)\citenamefont {Li},
  \citenamefont {Dong}, \citenamefont {Longhi}, \citenamefont {Liang},
  \citenamefont {Xie},\ and\ \citenamefont {Yan}}]{ABcage2022}%
  \BibitemOpen
  \bibfield  {author} {\bibinfo {author} {\bibfnamefont {H.}~\bibnamefont
  {Li}}, \bibinfo {author} {\bibfnamefont {Z.}~\bibnamefont {Dong}}, \bibinfo
  {author} {\bibfnamefont {S.}~\bibnamefont {Longhi}}, \bibinfo {author}
  {\bibfnamefont {Q.}~\bibnamefont {Liang}}, \bibinfo {author} {\bibfnamefont
  {D.}~\bibnamefont {Xie}},\ and\ \bibinfo {author} {\bibfnamefont
  {B.}~\bibnamefont {Yan}},\ }\href
  {https://doi.org/10.1103/PhysRevLett.129.220403} {\bibfield  {journal}
  {\bibinfo  {journal} {Phys. Rev. Lett.}\ }\textbf {\bibinfo {volume} {129}},\
  \bibinfo {pages} {220403} (\bibinfo {year} {2022})}\BibitemShut {NoStop}%
\bibitem [{\citenamefont {Raitzsch}\ \emph {et~al.}(2008)\citenamefont
  {Raitzsch}, \citenamefont {Bendkowsky}, \citenamefont {Heidemann},
  \citenamefont {Butscher}, \citenamefont {L\"ow},\ and\ \citenamefont
  {Pfau}}]{Raitzsch2008}%
  \BibitemOpen
  \bibfield  {author} {\bibinfo {author} {\bibfnamefont {U.}~\bibnamefont
  {Raitzsch}}, \bibinfo {author} {\bibfnamefont {V.}~\bibnamefont
  {Bendkowsky}}, \bibinfo {author} {\bibfnamefont {R.}~\bibnamefont
  {Heidemann}}, \bibinfo {author} {\bibfnamefont {B.}~\bibnamefont {Butscher}},
  \bibinfo {author} {\bibfnamefont {R.}~\bibnamefont {L\"ow}},\ and\ \bibinfo
  {author} {\bibfnamefont {T.}~\bibnamefont {Pfau}},\ }\href
  {https://doi.org/10.1103/PhysRevLett.100.013002} {\bibfield  {journal}
  {\bibinfo  {journal} {Phys. Rev. Lett.}\ }\textbf {\bibinfo {volume} {100}},\
  \bibinfo {pages} {013002} (\bibinfo {year} {2008})}\BibitemShut {NoStop}%
\bibitem [{\citenamefont {Yan}\ \emph {et~al.}(2020)\citenamefont {Yan},
  \citenamefont {Cincio},\ and\ \citenamefont
  {Zurek}}]{PhysRevLett.124.160603}%
  \BibitemOpen
  \bibfield  {author} {\bibinfo {author} {\bibfnamefont {B.}~\bibnamefont
  {Yan}}, \bibinfo {author} {\bibfnamefont {L.}~\bibnamefont {Cincio}},\ and\
  \bibinfo {author} {\bibfnamefont {W.~H.}\ \bibnamefont {Zurek}},\ }\href
  {https://doi.org/10.1103/PhysRevLett.124.160603} {\bibfield  {journal}
  {\bibinfo  {journal} {Phys. Rev. Lett.}\ }\textbf {\bibinfo {volume} {124}},\
  \bibinfo {pages} {160603} (\bibinfo {year} {2020})}\BibitemShut {NoStop}%
\bibitem [{\citenamefont {An}\ \emph {et~al.}(2021)\citenamefont {An},
  \citenamefont {Padavi\ifmmode~\acute{c}\else \'{c}\fi{}}, \citenamefont
  {Meier}, \citenamefont {Hegde}, \citenamefont {Ganeshan}, \citenamefont
  {Pixley}, \citenamefont {Vishveshwara},\ and\ \citenamefont
  {Gadway}}]{An2021}%
  \BibitemOpen
  \bibfield  {author} {\bibinfo {author} {\bibfnamefont {F.~A.}\ \bibnamefont
  {An}}, \bibinfo {author} {\bibfnamefont {K.}~\bibnamefont
  {Padavi\ifmmode~\acute{c}\else \'{c}\fi{}}}, \bibinfo {author} {\bibfnamefont
  {E.~J.}\ \bibnamefont {Meier}}, \bibinfo {author} {\bibfnamefont
  {S.}~\bibnamefont {Hegde}}, \bibinfo {author} {\bibfnamefont
  {S.}~\bibnamefont {Ganeshan}}, \bibinfo {author} {\bibfnamefont {J.~H.}\
  \bibnamefont {Pixley}}, \bibinfo {author} {\bibfnamefont {S.}~\bibnamefont
  {Vishveshwara}},\ and\ \bibinfo {author} {\bibfnamefont {B.}~\bibnamefont
  {Gadway}},\ }\href {https://doi.org/10.1103/PhysRevLett.126.040603}
  {\bibfield  {journal} {\bibinfo  {journal} {Phys. Rev. Lett.}\ }\textbf
  {\bibinfo {volume} {126}},\ \bibinfo {pages} {040603} (\bibinfo {year}
  {2021})}\BibitemShut {NoStop}%
\bibitem [{\citenamefont {Yao}\ \emph {et~al.}(2019)\citenamefont {Yao},
  \citenamefont {Khoudli}, \citenamefont {Bresque},\ and\ \citenamefont
  {Sanchez-Palencia}}]{PhysRevLett.123.070405}%
  \BibitemOpen
  \bibfield  {author} {\bibinfo {author} {\bibfnamefont {H.}~\bibnamefont
  {Yao}}, \bibinfo {author} {\bibfnamefont {H.}~\bibnamefont {Khoudli}},
  \bibinfo {author} {\bibfnamefont {L.}~\bibnamefont {Bresque}},\ and\ \bibinfo
  {author} {\bibfnamefont {L.}~\bibnamefont {Sanchez-Palencia}},\ }\href
  {https://doi.org/10.1103/PhysRevLett.123.070405} {\bibfield  {journal}
  {\bibinfo  {journal} {Phys. Rev. Lett.}\ }\textbf {\bibinfo {volume} {123}},\
  \bibinfo {pages} {070405} (\bibinfo {year} {2019})}\BibitemShut {NoStop}%
\bibitem [{\citenamefont {Liu}\ \emph {et~al.}(2015)\citenamefont {Liu},
  \citenamefont {Ghosh},\ and\ \citenamefont {Chong}}]{PhysRevB.91.014108}%
  \BibitemOpen
  \bibfield  {author} {\bibinfo {author} {\bibfnamefont {F.}~\bibnamefont
  {Liu}}, \bibinfo {author} {\bibfnamefont {S.}~\bibnamefont {Ghosh}},\ and\
  \bibinfo {author} {\bibfnamefont {Y.~D.}\ \bibnamefont {Chong}},\ }\href
  {https://doi.org/10.1103/PhysRevB.91.014108} {\bibfield  {journal} {\bibinfo
  {journal} {Phys. Rev. B}\ }\textbf {\bibinfo {volume} {91}},\ \bibinfo
  {pages} {014108} (\bibinfo {year} {2015})}\BibitemShut {NoStop}%
\bibitem [{\citenamefont {Iyer}\ \emph {et~al.}(2013)\citenamefont {Iyer},
  \citenamefont {Oganesyan}, \citenamefont {Refael},\ and\ \citenamefont
  {Huse}}]{PhysRevB.87.134202}%
  \BibitemOpen
  \bibfield  {author} {\bibinfo {author} {\bibfnamefont {S.}~\bibnamefont
  {Iyer}}, \bibinfo {author} {\bibfnamefont {V.}~\bibnamefont {Oganesyan}},
  \bibinfo {author} {\bibfnamefont {G.}~\bibnamefont {Refael}},\ and\ \bibinfo
  {author} {\bibfnamefont {D.~A.}\ \bibnamefont {Huse}},\ }\href
  {https://doi.org/10.1103/PhysRevB.87.134202} {\bibfield  {journal} {\bibinfo
  {journal} {Phys. Rev. B}\ }\textbf {\bibinfo {volume} {87}},\ \bibinfo
  {pages} {134202} (\bibinfo {year} {2013})}\BibitemShut {NoStop}%
\bibitem [{\citenamefont {Lyubarov}\ \emph {et~al.}(2022)\citenamefont
  {Lyubarov}, \citenamefont {Lumer}, \citenamefont {Dikopoltsev}, \citenamefont
  {Lustig}, \citenamefont {Sharabi},\ and\ \citenamefont
  {Segev}}]{Lyubarov2022}%
  \BibitemOpen
  \bibfield  {author} {\bibinfo {author} {\bibfnamefont {M.}~\bibnamefont
  {Lyubarov}}, \bibinfo {author} {\bibfnamefont {Y.}~\bibnamefont {Lumer}},
  \bibinfo {author} {\bibfnamefont {A.}~\bibnamefont {Dikopoltsev}}, \bibinfo
  {author} {\bibfnamefont {E.}~\bibnamefont {Lustig}}, \bibinfo {author}
  {\bibfnamefont {Y.}~\bibnamefont {Sharabi}},\ and\ \bibinfo {author}
  {\bibfnamefont {M.}~\bibnamefont {Segev}},\ }\href
  {https://doi.org/10.1126/science.abo3324} {\bibfield  {journal} {\bibinfo
  {journal} {Science}\ }\textbf {\bibinfo {volume} {377}},\ \bibinfo {pages}
  {425} (\bibinfo {year} {2022})}\BibitemShut {NoStop}%
\bibitem [{\citenamefont {Heugel}\ \emph {et~al.}(2019)\citenamefont {Heugel},
  \citenamefont {Oscity}, \citenamefont {Eichler}, \citenamefont {Zilberberg},\
  and\ \citenamefont {Chitra}}]{Heugel2019}%
  \BibitemOpen
  \bibfield  {author} {\bibinfo {author} {\bibfnamefont {T.~L.}\ \bibnamefont
  {Heugel}}, \bibinfo {author} {\bibfnamefont {M.}~\bibnamefont {Oscity}},
  \bibinfo {author} {\bibfnamefont {A.}~\bibnamefont {Eichler}}, \bibinfo
  {author} {\bibfnamefont {O.}~\bibnamefont {Zilberberg}},\ and\ \bibinfo
  {author} {\bibfnamefont {R.}~\bibnamefont {Chitra}},\ }\href
  {https://doi.org/10.1103/PhysRevLett.123.124301} {\bibfield  {journal}
  {\bibinfo  {journal} {Phys. Rev. Lett.}\ }\textbf {\bibinfo {volume} {123}},\
  \bibinfo {pages} {124301} (\bibinfo {year} {2019})}\BibitemShut {NoStop}%
\end{thebibliography}%

\end{document}